\pdfoutput=1
\documentclass[conference]{IEEEtran}
\usepackage[pdftex]{graphicx}
\usepackage[cmex10]{amsmath}
\usepackage{url}
\usepackage{arabtex}
\usepackage{verbatim}
\usepackage{fancyhdr}
\usepackage{hyperref}
\fancypagestyle{firstpage}{% Page style for first page
  \fancyhf{}% Clear header/footer
  
  \fancyfoot[L]{{\em This paper was published in the Proceedings of the 2016 IEEE Tenth International 
Conference on Semantic Computing (ICSC), pages 79-86, Laguna Hills, CA, USA, February 2016. \copyright 2016 IEEE\\
Link to article abstract in IEEE Xplore: http://dx.doi.org/10.1109/ICSC.2016.38}}% Footer
}
\begin{document}

\pagenumbering{gobble}

\title{Data Cleaning for XML Electronic Dictionaries via Statistical Anomaly Detection}

\IEEEoverridecommandlockouts

\author{\IEEEauthorblockN{Michael Bloodgood}
\IEEEauthorblockA{Center for Advanced Study of Language\\
University of Maryland\\
College Park, MD 20742\\
meb@umd.edu}
\and
\IEEEauthorblockN{Benjamin Strauss\dag \thanks{\dag This research was conducted while this author was a Faculty Research Assistant at the University of Maryland Center for Advanced Study of Language.}}
\IEEEauthorblockA{Department of Computer Science and Engineering\\
The Ohio State University\\
Columbus, OH 43210\\
strauss.105@osu.edu}
}

\maketitle

\thispagestyle{firstpage}% firstpage page style for first page

\begin{abstract}
Many important forms of data are stored digitally in XML format. Errors can occur in the textual content of the data in the fields of the XML. 
Fixing these errors manually is time-consuming and expensive, especially for large amounts of data. There is increasing interest in 
the research, development, and use of automated techniques for assisting with data cleaning. Electronic dictionaries are an important form of 
data frequently stored in XML format that frequently have errors introduced through a mixture of manual typographical entry errors and optical character recognition errors. 
In this paper we describe methods for flagging statistical anomalies as likely errors in electronic dictionaries stored in XML format. 
We describe six systems based on different sources of information. The systems detect errors using various signals in the data including 
uncommon characters, text length, character-based language models, word-based language models, tied-field length ratios, and tied-field 
transliteration models. Four of the systems detect errors based on expectations automatically inferred from content within elements of a single field type. 
We call these {\em single-field} systems. 
Two of the systems detect errors based on correspondence expectations automatically inferred from content within elements of multiple related field types. 
We call these {\em tied-field} systems.
For each system, we provide an intuitive analysis of the type of error that it is successful at detecting. 
Finally, we describe two larger-scale evaluations using crowdsourcing with Amazon's Mechanical Turk platform and using the annotations of a domain expert.
The evaluations consistently show that the systems are useful for improving the efficiency with which errors in 
XML electronic dictionaries can be detected. 
\end{abstract}

\IEEEpeerreviewmaketitle

\section{Introduction} \label{introduction}

There is increasing interest in the research, development, and use of automated techniques for assisting with {\em data cleaning}, also 
called {\em data cleansing} or {\em scrubbing}, which deals with detecting and removing errors and inconsistencies from data in order to improve the 
quality of data \cite{rahm2000}. In this paper we deal with data cleaning of electronic dictionaries stored in Extensible Markup Language (XML) format. XML is a markup 
language that defines a set of rules for encoding documents in a format that is both human-readable and machine-readable. Defined by free open 
standards, XML is a textual data format with strong support via Unicode for different human languages. It is widely used for the representation of 
electronic dictionaries and other forms of structured data \cite{ide1995, francopoulo2006}.

Electronic dictionaries are a fundamentally important semantic computing resource. They are a core resource consumed by downstream processes in the 
provision of various human language technologies as well as consumed directly by human language learners and as reference materials more generally. 
When dictionaries are digitized, whether via manual entry, Optical Character Recognition (OCR), or a mixture of these methods, it is inevitable 
that errors are introduced into the digitized version that is produced. 

Prior work has focused on providing editing tools to assist with manual 
curation of the data and on providing tools for automatically detecting structural errors using only structural information. 
Although these systems have had success in finding dictionary errors, there are many errors that cannot be detected without analyzing the text content of the dictionary. 
For example, suppose that in some dictionary a lexical entry requires a headword, a part of speech, and a definition. 
Suppose there is an error in which the definition for some entry appears in the part of speech field, and the definition field contains the headword of the next entry.
These errors will not be found by examining only the structure. 
Unlike previous work, in this paper we present methods for flagging statistical anomalies as likely errors in the textual content itself in electronic 
dictionaries in XML format using information from the textual content of the data. 

We present six systems that detect errors using various signals in the data. The types of data quality problems that our systems are designed to detect 
are single-source instance level data quality problems \cite{rahm2000}. 
Four of our systems detect errors based on information automatically inferred from content within elements of a single field type. 
We call these systems {\em single-field} systems. 
Two of the systems detect errors based on correspondence information automatically inferred from content within elements of multiple related field types. 
We call these latter systems {\em tied-field} systems. 
For each system, we provide an intuitive analysis of the type of error that it is successful at detecting. 
Finally, we describe two larger-scale evaluations using crowdsourcing with Amazon's Mechanical Turk platform and using domain expert annotations.
The evaluations consistently show that the systems are useful for improving the efficiency with which errors in XML electronic dictionaries can be detected. 

In the next section we situate our work with respect to previous related work. In section~\ref{methods} we describe the error detection systems in 
detail and provide intuitive examples of the sorts of errors that each error detection system finds. In section~\ref{evaluation} we describe our 
experimental tests and provide experimental results and discussion of results. Finally, in section~\ref{conclusion} we conclude. 

\section{Related Work} \label{relatedWork}

A categorization of data quality problems addressed by data cleaning and an overview of data cleaning methods is provided in \cite{rahm2000}. 
Many data cleaning methods are based on identifying discrepancies for user auditing, e.g., \cite{raman2001,guyon1996}. 
The system in \cite{raman2001} is highly interactive; discrepancy detection is not their sole main focus. 
The discrepancy detection approach they outline is for users to define domains and then for 
discrepancies to be located through checking against the user-defined domains for constraint violation. 
In contrast, the methods in the current paper do not require user specification of domains and the methods in the current paper operate 
on the basis of different sources and indicators of discrepancies. 
In particular, the methods in the current paper are fundamentally different in that they operate on the basis of statistically anomalous events instead of 
constraint violations. 
In \cite{guyon1996}, a similar overall workflow is presented whereby the most suspicious examples are flagged for a human operator to inspect and 
annotate as clean or ``garbage." 
In contrast to the current paper, there is only one method in \cite{guyon1996}, which is to flag the examples with the highest information gain for 
inspection by a human operator. 
The work in \cite{guyon1996} assumes the context of construction of an automated classifier in the computation of the information gain. 
The method in \cite{guyon1996} is evaluated on the task of handwritten digit recognition by seeing how well classifier performance improves with the 
addition of the data cleaning approach. 
In contrast, the current paper uses different methods for detecting suspicious examples and does not assume the context of construction of automated classifiers. 
Also in contrast, the methods in the current paper are evaluated on XML electronic dictionaries by measuring how many and what percentages of the 
detected anomalies are annotated as data errors by domain experts. 
The methods in the current paper may be able to be used in a complementary fashion with the methods from \cite{raman2001} and \cite{guyon1996} in future work. 

Past work presented a method for repairing a digital dictionary in an XML format using a dictionary markup language called DML \cite{zajic2011}.
It remains time-consuming and error-prone however to have a human exhaustively read through and manually correct a digital version of a dictionary, even 
with languages such as DML available for making corrections once errors are detected. The methods we present in the current paper automatically scan through and
detect errors in dictionaries. The methods in the current paper can be used in concert with error correction techniques such as dictionary markup languages. 

Previous approaches have been presented for detecting structural errors in digitized dictionaries \cite{bloodgood2012, rodrigues2011}. 
The method in \cite{rodrigues2011} works by linearizing the lexicon structure, converting the opening tags in XML into tokens and then considering the likelihoods of
various strings of tokens using a language modeling approach. Anomalous branches of XML tags are flagged as structural errors. The method ignores the 
underlying text data within the dictionary and only detects structural errors. The methods in \cite{bloodgood2012} outperform the method from \cite{rodrigues2011}. 
The methods in \cite{bloodgood2012} use a mixture of unsupervised methods, supervised machine learning methods, and system combination
approaches. The highest-performing method uses a random forest system combination approach. The methods in \cite{bloodgood2012}
only detect structural errors. Errors in the textual data content within the XML elements are not detected. 
In contrast, the current paper presents methods that detect errors in the text (data) content of XML elements. 
The methods in the current paper also use different approaches and different sources of
information than were used in \cite{bloodgood2012, rodrigues2011} and do not require training data, which is often not available. 
The error detection methods in the current paper can be used in concert with 
structural error detection methods.   

\section{Methods} \label{methods}

This section describes how our methods work. 
We categorize our methods into two types: {\em single-field} methods and {\em tied-field} methods.
Single-field methods work by utilizing information within the data content of a single XML field type in order to detect errors. 
Tied-field methods work by utilizing information within the data content of multiple XML field types, exploiting various relationships between the data in the different
fields, in order to detect errors. 

For each candidate error, all of our methods return a numeric score indicating the system's confidence that the candidate is an error. 
A threshold can be set for each method to control which candidates are detected as errors. 
The threshold for each method can be adjusted according to recall-precision\footnote{Recall and precision are standard measures for systems that perform search. 
For the case of detecting dictionary errors, recall is the percentage of true errors that are found by the system. Precision is the percentage of system-proposed errors
that are in fact true errors in the dictionary.} preferences and dictionary characteristics. 
In general, higher thresholds will return results with higher precision and lower recall whereas lower thresholds will return results with lower precision and higher recall.
The optimal threshold for each method depends on data characteristics and user preferences. 
We are not aware of a method for determining optimal thresholds.
We set our thresholds to a level that yielded a reasonable number of error candidates for human review. The exact thresholds for each experiment are given in the following subsections.

Subsection~\ref{single-FieldMethods} describes how the single-field methods work and 
subsection~\ref{tied-FieldMethods} describes how the tied-field methods work. 
Subsection~\ref{examples} provides examples of anomalies detected by the various methods.

\subsection{Single-Field Methods} \label{single-FieldMethods}

The single-field error detection systems do not require any advance knowledge about the structure of the
electronic dictionary. The only structural context information they use is the tag name of the elements containing
the text data to be checked. Single-field methods can be used to check for errors in elements of any individual tag name. 
All single-field methods work according to the following high-level description: all entries of an 
individual tag name are processed and then any entries that are anomalous are flagged as errors. Each single-field method processes the entries and flags anomalies in
different ways based on different aspects of the data. The rest of this subsection describes four single-field methods in detail. 

\subsubsection{Uncommon Characters Method} \label{uncommonCharactersMethod}

Uncommon characters can be a frequent source of errors in electronic dictionaries. They can arise due to
OCR errors, author typographical errors, and mislabeled and/or incorrectly
merged fields. Texts that contain uncommon characters are reported as potential errors. For each element in the
dictionary with a particular tag, we consider the texts inside those elements to be a collection of documents $D$,
and the characters in the texts as the tokens. We calculate the inverse document frequency of each character $c$
observed in $D$ as follows:
\begin{equation} \label{idf}
idf(c,D) = \log_{10} \frac{N}{|\{ d \in D : c \in d \}|},
\end{equation}
where $N$ is the number of documents in the collection.
 
When $idf(c,D) > threshold$, we consider $c$ to be an uncommon character.   
The threshold is configurable; we use a default value of four. 
Users can adjust the threshold according to their recall-precision preferences and according to dictionary characteristics.  
The elements containing uncommon characters are flagged by the system as potential errors.

\subsubsection{Text Length Method} \label{textLengthMethod}

When two text fields are inappropriately combined into one field, the result can be text that is unusually
long. When a single text unit is inappropriately truncated, or split across two fields, the result can be text that is
unusually short. Texts with unusually long or unusually short length are reported as potential errors. For each
element in the dictionary with a given tag, we treat the lengths of the texts inside those elements as a sample of a
normally distributed population. We calculate the mean and standard deviation of the sample, and for each
value, we calculate the z-score, i.e., the signed number of standard deviations the value is above or below the
mean. If the absolute value of the z-score of the length of a text is above a threshold, we flag the text as a possible error. 
The threshold is configurable; we use a default value of four. As the threshold is raised, only the most unusually long, or short, fields will be returned
as errors. 

\subsubsection{Word Sequence Method} \label{wordSequenceMethod}

Language modeling can capture the probability of a sequence of words occurring in textual content. This
gives us the capability to flag unlikely sequences of words. These unlikely sequences can often be indicative of
typographical errors, OCR errors, incorrect field joining and splitting, etc. Texts that are unlikely given a
word-level language model of texts in the same context are reported as potential errors.
For each element in the dictionary with a given tag, we build a language model of the text content of all the elements using
n-grams of words. We calculate the entropy of the model with respect to each individual element's text, and treat the entropies as a
sample of a normally distributed population. We calculate the z-score of each entropy score. High z-scores
indicate texts that are unlikely in the language model. The texts with entropy z-scores above a threshold are
flagged as errors. The size of the n-grams in the language model and the threshold are configurable. 
We use 4-grams and threshold five by default. Using a larger n-gram size in the language model allows one to potentially
capture more nuanced sequence characteristics, however, it would require a much larger amount of data to
estimate properly and avoid introducing spurious estimates due to data sparsity. Also, larger n-gram sizes are
more computationally intensive. The entropy z-score thresholds can be adjusted according to recall-precision
preferences and dictionary characteristics. 

\subsubsection{Character Sequence Method} \label{characterSequenceMethod}

This method is similar to the Word Sequence
Method, except that instead of n-grams of words, we build language models using n-grams of characters. The
size of the n-grams in the language model and the threshold are configurable. We use 4-grams and a threshold of five
by default. Larger n-gram sizes could potentially capture more nuanced character sequence models, however,
they would require a much larger amount of data to estimate properly and avoid introducing spurious estimates
due to data sparsity. The n-gram size can also be adjusted based on the language of the field's textual content.
The entropy z-score thresholds can be adjusted according to recall-precision preferences and dictionary
characteristics. 
Note that the error detection system based on character sequences will in some cases find errors that the Uncommon
Character system also detects, but the Character Sequence Method is also capable of finding some errors that the Uncommon Character system is not
able to find. This is because the Character Sequence Method can find errors in which none of the characters in the
textual content is particularly uncommon, but in which the ordering of those characters is incorrect.

\subsection{Tied-Field Methods} \label{tied-FieldMethods}

In many structured data sets, there are pairs of fields that are related to each other in predictable ways. 
For example, in dictionaries a word in a language's native orthography and a phonetic transcription of the word are related because in many languages 
there are predictable relationships between spelling and pronunciation. 
Another related pair of fields in bilingual dictionaries is an example sentence demonstrating usage of a word and its translation. 
We call these sorts of related fields {\em tied fields} and we call error detection methods that exploit relationships between content in 
different field types {\em tied-field} methods. 
It is possible for there to be errors in a single field where the data value in that field is not anomalous in the 
context of only other values in that field type. 
However, the value might be anomalous in the context of data values in related fields. 
The single-field methods presented in subsection~\ref{single-FieldMethods} will be unable to detect these sorts of errors. 
The rest of this subsection describes two tied-field methods that can detect these sorts of errors. 

\subsubsection{Tied-Field Length Ratio Method} \label{tied-FieldLengthRatioMethod}

This method determines the ratio of length in characters of tied fields; pairs of data values with unusual length ratios are then reported as potential errors. 
We treat the ratio of the length in characters of the tied-field data values to be a sample of a normally distributed population. 
We calculate the absolute value of the z-score of each length ratio. 
High values indicate tied-field instances where the data values have an unusual length ratio.
Tied-field instances with scores above a threshold are flagged as potential errors. 
The threshold is configurable; we use a threshold of two by default. 
The threshold can be adjusted according to recall-precision preferences and dictionary characteristics. 

To handle situations in which the distribution of length ratios is significantly different for short and long
strings, we have an option to partition the tied-field instances by the length of the data in the first field, and treat each partition as its own population. 
Table~\ref{t:shortAndLongStrings} illustrates why it could be beneficial to partition tied-field instances by length. In these examples of
correct pairs of data, the length ratios of the short tied-field pairs of data can be seen to vary more than 
the length ratios of the long tied-field pairs of data. 

\begin{table}
\caption{Examples of orthography-pronunciation pairs. Short strings have a different distribution of length ratios than long strings.}
\label{t:shortAndLongStrings}
\centering
\begin{tabular}{|c|c|c|c|c|c|}
\hline
\multicolumn{3}{|c|}{Short} & \multicolumn{3}{|c|}{Long} \\
\hline
Orth & Pron & Ratio & Orth & Pron & Ratio \\
\hline
t & t\={e} & 0.50 & groundwork & ground"w\^{u}rk` & 0.83 \\
\hline
ease & \={e}z & 2.00 & lithargyrum & l\u{i}*th\"{a}r"j\u{i}*r\u{u}m & 0.79 \\
\hline
v & v\={e} & 0.50 & haidingerite & h\={i}"d\u{i}ng*\~{e}r*\={i}t & 0.92 \\
\hline
\end{tabular}
\end{table}

\subsubsection{Tied-Field Transliteration Method} \label{tied-FieldTransliterationMethod}

For some tied fields, the data in the two fields have a more specific relationship than a length relationship. 
For example, in dictionaries some fields are transliterations of each other. 
Such transliterations will usually have character-level correspondences (not
necessarily a one-to-one correspondence). If the correspondence can be modeled, then pairs of texts that
do not correspond well can be reported as errors. 
We use Phonetisaurus\footnote{\url{https://code.google.com/p/phonetisaurus/}} to learn transliteration models across tied-fields. 
The Phonetisaurus
system has been described in detail and has been shown to perform well in \cite{novak2012}. 
The resulting transliteration model represents how the first field can be transliterated into the second field, and can be used to generate
scored transliteration candidates of the first field.
For each pair of tied-fields, we use the transliteration model to transliterate the first field into the n-best
candidates for the second field. 
Each candidate is given a transliteration cost by Phonetisaurus. 
By taking the inverse of the cost, we obtain a score indicating the model's confidence in that transliteration candidate. 
We calculate the normalized edit distance (NED) of each candidate to the observed data that actually is present in the second field, and calculate the
mean NED weighted by transliteration score.
We treat the weighted means as a sample of a normally distributed population. We calculate the z-score of
each weighted mean. High values indicate that the data occurring in the second field is an unusually large NED away from what our learned model would have expected based 
on the data that occurred in the first field. 
Instances with weighted mean z-scores above a threshold are flagged as errors. 
The threshold is configurable; we use a default of two. 
The weighted mean z-score threshold can be adjusted according to recall-precision preferences and dictionary characteristics.\footnote{We also offer an option to normalize the tied-field text data to all lowercase letters.} 

\subsection{Examples} \label{examples}

In this subsection we provide examples of anomalies detected by the various systems. 
To obtain these examples, we ran the systems over a digitized sample of an Urdu-English dictionary \cite{qureshi1991} and selected 
illustrative examples that can be understood by most readers without having to know too many details about specific dictionary representations in XML. 
Table~\ref{t:examplesSingleField} shows examples detected by the various single-field systems and Table~\ref{t:examplesTiedField} shows an 
example detected by the tied-field systems. 

\begin{table}
\caption{Examples of anomalies detected by single-field systems.}
\label{t:examplesSingleField}
\centering
\begin{tabular}{|c|c|c|}
\hline
ExampleID & FieldName & Value \\
\hline
Example 1 &  GENDER & /F. \\
\hline
Example 2 & NUMBER & PLU. \\
\hline
Example 3 & PART-OF-SPEECH & PARTICLE \\
\hline
\end{tabular}
\end{table}

\begin{table}
\caption{Example anomaly detected by tied-field systems.}
\label{t:examplesTiedField}
\centering
\begin{tabular}{|c|c|c|c|c|}
\hline
ExampleID & FieldName & Value \\
\hline
Example 4 &  ORTHOGRAPHY & \<رجعت قہقری> \\ \cline{2-3}
& PRONUNCIATION &  r\={a} \\
\hline
\end{tabular}
\end{table}

Example 1 in Table~\ref{t:examplesSingleField} was detected by all four of our single-field systems. 
This is an error in the data since the value should have been just ``F.'' without the ``/''. 
Since the GENDER field almost exclusively contains values of ``M.'' or ``F.'', the Word Sequence Method found any other words such as ``/F.'' to have 
unusually high entropy. The character-based language model detected this error similarly. 
The Text Length Method detected this error since it is three characters long, which is unusually long 
given the predominance of two-character-long values in this field. 
The Uncommon Characters Method detected this error since the ``/'' character is uncommon in this field. 
Example 1 shows how the different methods can sometimes all detect the same error albeit through different views of the data.

Example 2 in Table~\ref{t:examplesSingleField} shows an example of an error that was detected by some of the systems and not others. 
This is an error in the data since the value should have been just ``PL.'' without the ``U''. The Word Sequence Method, the Character Sequence Method, and the Uncommon Characters Method all found this error. The Uncommon Characters Method detected this error since ``U'' is an uncommon character for values in the NUMBER field. 
The Text Length Method did not detect this error. This is because the length of four characters is not unusually long or short for this field - the NUMBER field often contains ``PL.'' with length three characters and ``SING.'' with length five characters. 

Example 3 in Table~\ref{t:examplesSingleField} shows an example of an anomaly that was detected by all four single-field methods that was not erroneous. 
The Word Sequence Method and the Character Sequence Method detected it since it had unusually high entropy. The Text Length Method detected it since it was 
unusually long. The Uncommon Characters Method detected it since the characters ``C", ``E", ``L", ``P", and ``R" are uncommon for this field. 
For reference, some of the most common values for the part of speech field include ``V.", ``N.", ``ADJ.", ``ADV.", etc. 
Although ``PARTICLE" is an uncommon value for this field, it is not an error.

Example 4 in Table~\ref{t:examplesTiedField} shows an example detected by both the Tied-Field Text Length Ratio Method and the Tied-Field Transliteration Method. 
The ORTHOGRAPHY field had the value ``\<رجعت قہقری>" and the PRONUNCIATION field had the value ``r\={a}". 
This is an error resulting from incorrect merging and splitting of fields.
The Text Length Ratio Method detected this as an error since the ratio was unusually high for values of these two fields. 
The Transliteration Method detected this as an error since the weighted mean Normalized Edit Distance from the generated pronunciation candidates to the observed pronunciation 
``r\={a}" was unusually high. Note that the single-field methods did not detect this error since ``\<رجعت قہقری>" is not an anomalous value for ORTHOGRAPHY in isolation and neither is ``r\={a}" an anomalous value for PRONUNCIATION in isolation. It is only when they are tied to each other that an anomaly is detected. 

The examples help to illustrate on a small scale what types of errors the various methods can detect. 
Also, the examples show how sometimes the 
methods have overlapping behavior and sometimes the methods have complementary behavior. 
The examples also illustrate how sometimes the methods 
detect errors in the data that need to be corrected and sometimes the methods detect anomalies that, while rare, are not errors that need to be corrected. 
In the next section we provide larger-scale evaluations of the error detection methods. 

\section{Evaluation} \label{evaluation}

We used Amazon's Mechanical Turk crowdsourcing platform to evaluate the Tied-Field Length Ratio Method, the Tied-Field Transliteration Method, and a 
random sample of data. Mechanical Turk is an online crowdsourcing platform
where workers, also called Turkers, complete simple tasks called Human Intelligence Tasks (HITs). Crowdsourcing can allow inexpensive and rapid data collection 
for various Natural Language Processing (NLP) tasks \cite{snow2008, bloodgood2010a, negri2011}, including human evaluations of NLP systems \cite{callison-burch2009, bloodgood2010b, amigo2010, chen2011, bloodgood2014}. 

The tied fields that we used in our evaluation were the orthography and the corresponding pronunciation fields from the GNU Collaborative International Dictionary of English (GCIDE).
GCIDE is a freely available dictionary of English based on Webster's 1913 Revised
Unabridged Dictionary and supplemented with entries from WordNet \cite{miller1990, miller1995, fellbaum1998} and additional submissions from users.  
GCIDE is formatted in XML and is available for download from \url{www.ibiblio.org/webster/}. Out of the 16704 pairs of orthography-pronunciation values in the dictionary, our tied-field error detection systems 
identified 2797 of the pairs as possible errors. From this set of detected candidate errors, we randomly selected 1000 pairs detected by the tied-field length ratio system and 1000 pairs detected by the tied-field transliteration system for evaluation by Turkers. 

\begin{figure}
\centering
\includegraphics[width=2.5in]{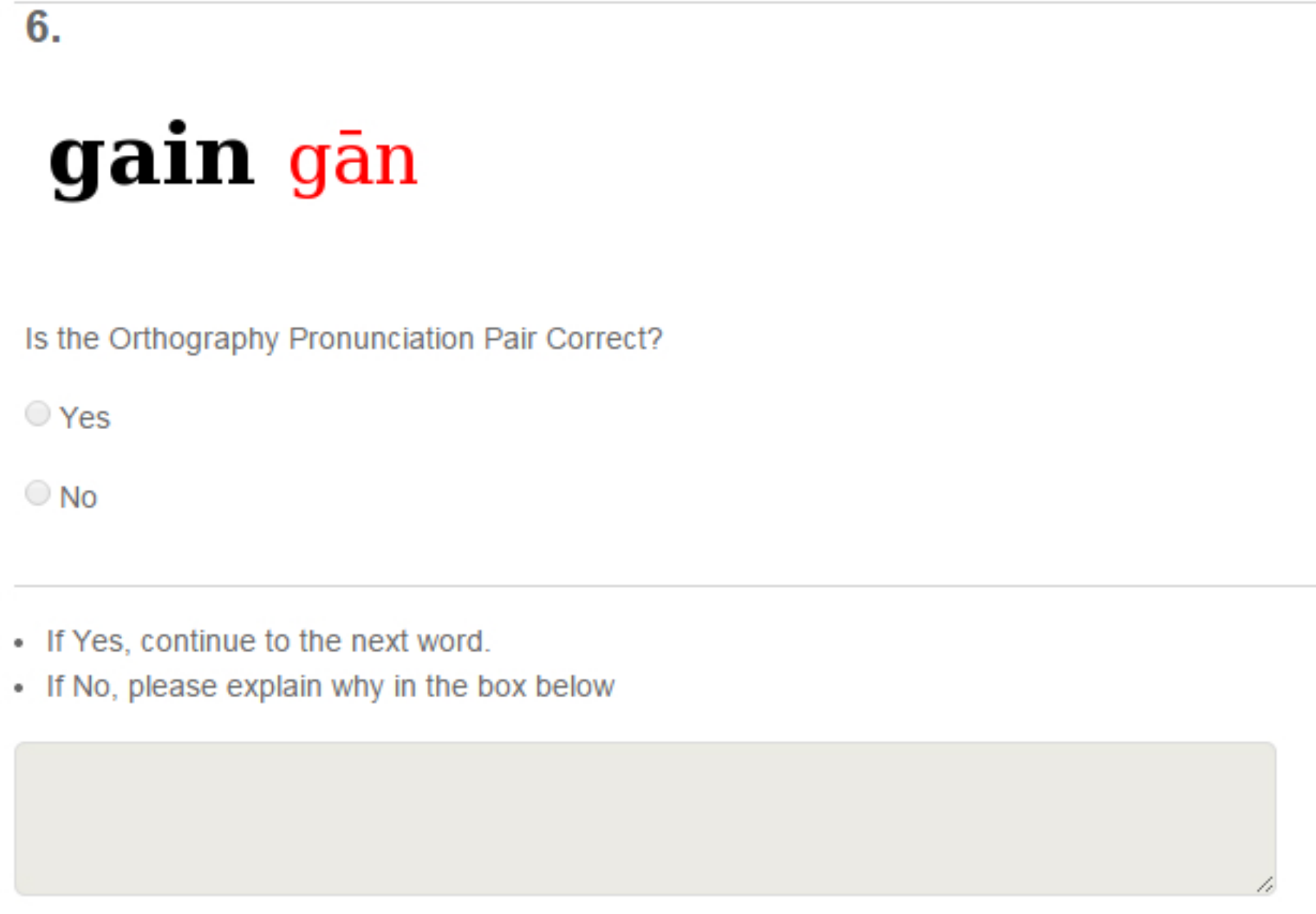}
\caption{Screenshot of interface for evaluating tied-field error detection systems.}
\label{f:screenshot}
\end{figure}

For each candidate error, we asked five Turkers if the orthography-pronunciation pair was correct. 
Figure~\ref{f:screenshot} shows a screenshot of the Mechanical Turk interface we used. 
If a Turker judged that a pair was not correct, i.e., that the pair was truly an error, then the Turker was required to provide an explanation. 
By requiring an explanation when pairs are incorrect, we are, if anything, creating a bias where workers will tend towards saying pairs are correct 
since that is easier for them. This will cause, if anything, the efficacy of our error detection systems to be understated. 

\begin{figure}
\centering
\includegraphics[width=2.5in]{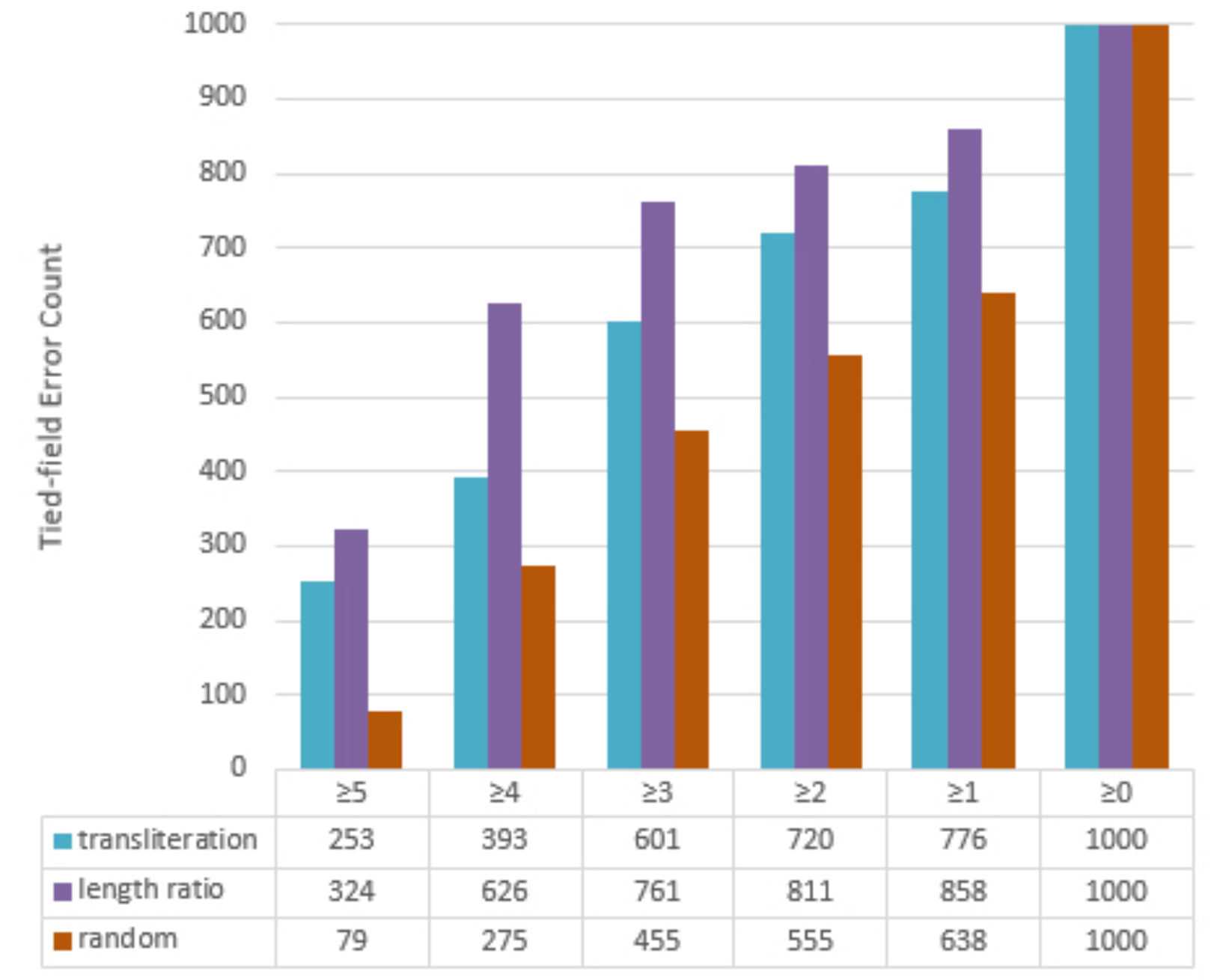}
\caption{Counts of proposed errors judged to be real errors by at least 5, 4, 3, 2, 1, and 0 Turkers.}
\label{f:counts}
\end{figure}

Figure~\ref{f:counts} displays counts of proposed orthography-to-pronunciation errors judged to be real errors by the
Turkers for the tied-field transliteration and the tied-field length ratio error detection systems as well as for
randomly selected examples. 
The x-axis shows the number of Turkers that agreed a proposed error was a real
error. Recall that for each of the proposed errors we had asked five Turkers to judge whether it was a real error
or not. The y-axis shows the number of proposed errors that were judged to be real errors. The main observation is that both the
tied-field transliteration system and the tied-field length ratio system find many more errors than
the random selection system. In particular, observe that for the tied-field transliteration system three or more
Turkers agree its proposed errors are really errors more than 60\% of the time; for the tied-field length ratio
system three or more Turkers agree its proposed errors are really errors more than 76\% of the time. In contrast,
for randomly selected proposed errors, three or more Turkers agree they are really errors only about 45.5\% of the time. 
These results are evidence that the error detection systems could substantially increase the efficiency with which domain experts can clean XML data.
 
\begin{figure}
\centering
\includegraphics[width=2.5in]{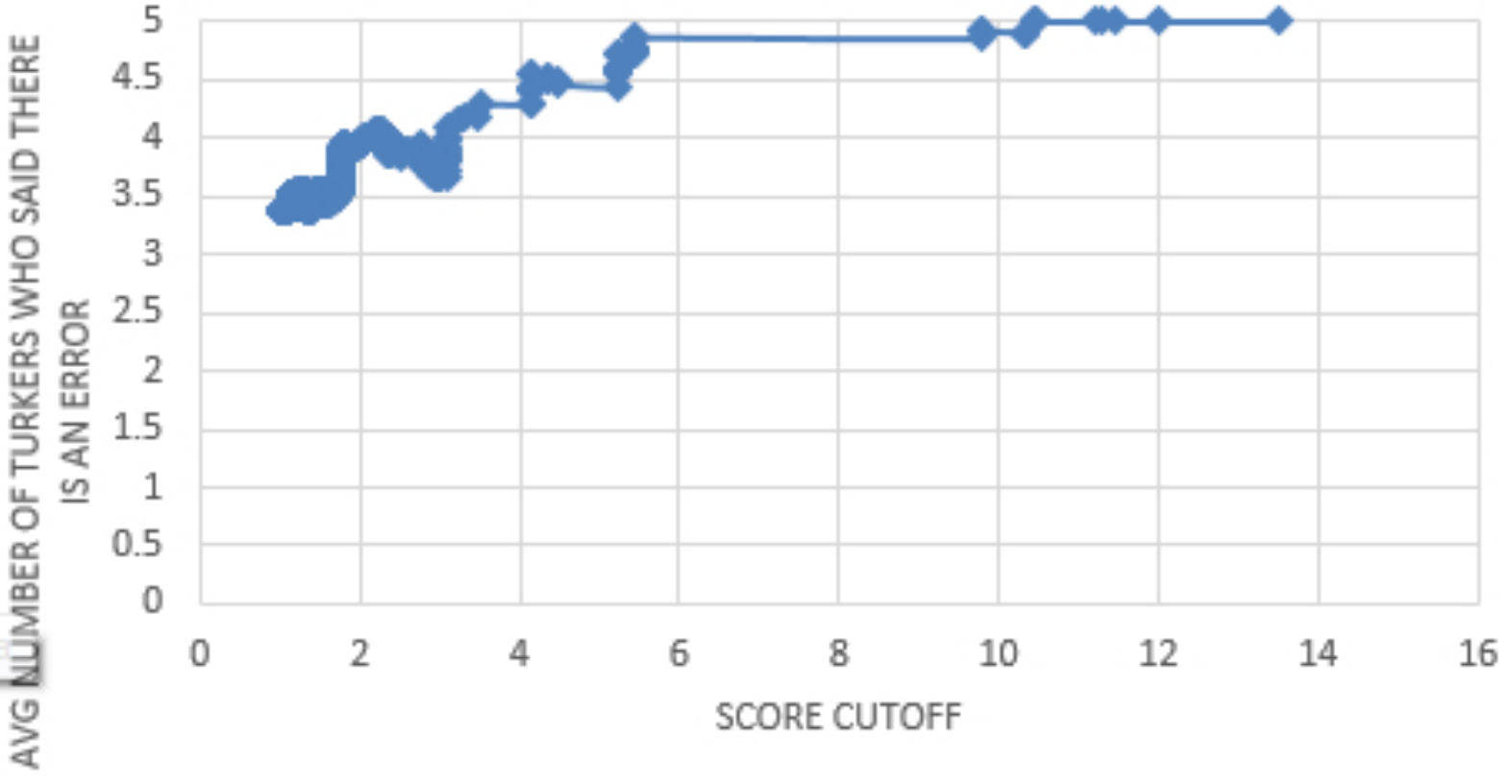}
\caption{The average number of Turkers (out of 5) who believe the orthography-pronunciation pair is an error for varying score cutoffs for 
the length ratio system. The score is the absolute value of the z-score.}
\label{f:lengthRatioScoreCutoff}
\end{figure}

\begin{figure}
\centering
\includegraphics[width=2.5in]{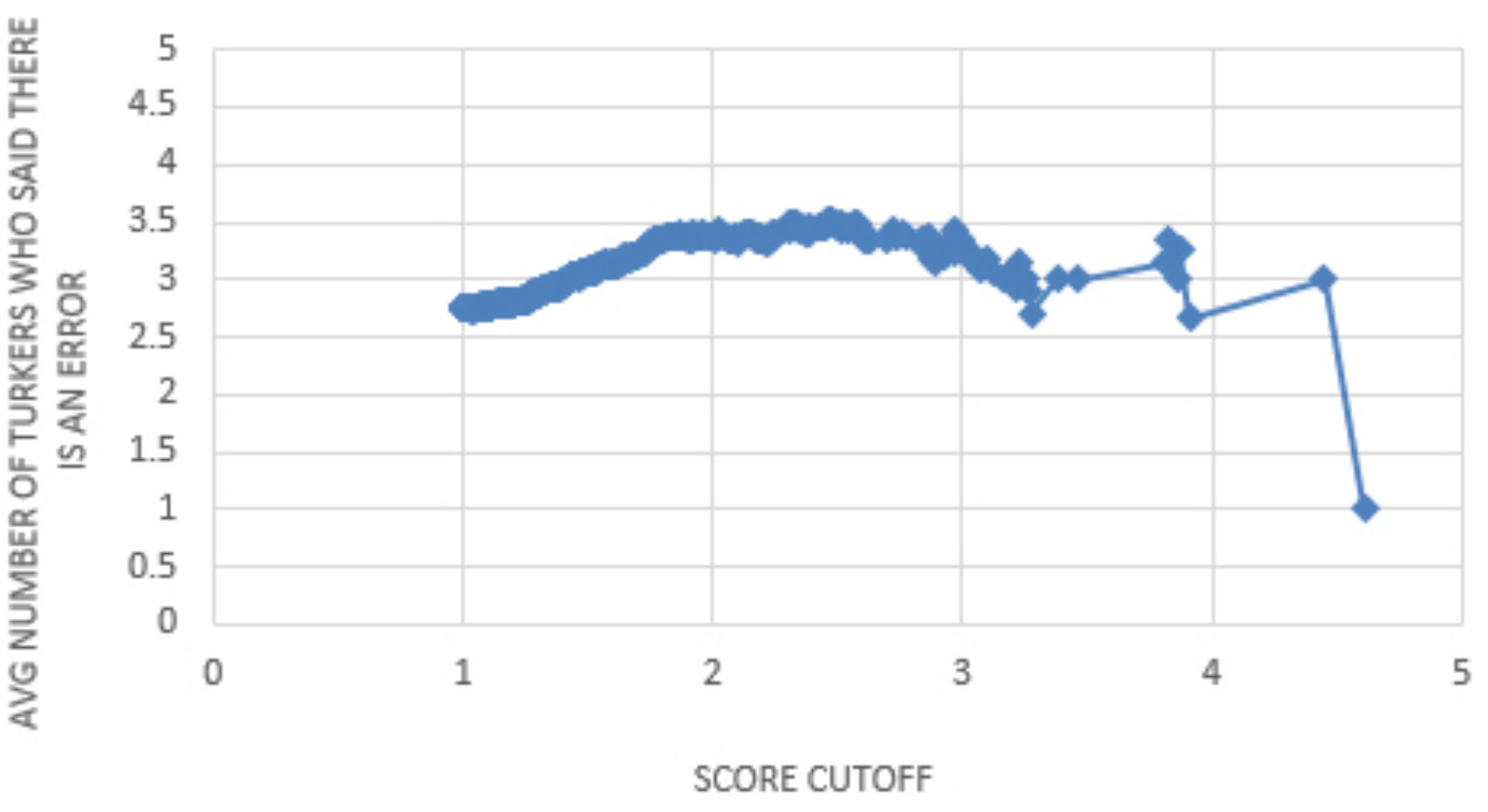}
\caption{The average number of Turkers (out of 5) who believe the orthography-pronunciation pair is an error for varying score cutoffs for 
the transliteration system. The score is the z-score.}
\label{f:transliterationScoreCutoff}
\end{figure}

Diving a little deeper, Figure~\ref{f:lengthRatioScoreCutoff} and Figure~\ref{f:transliterationScoreCutoff} show the average number of Turkers that agree a proposed error is really an error
for proposed errors at varying score cutoffs. Figure~\ref{f:lengthRatioScoreCutoff} shows the information for the tied-field length ratio error
detection system. In Figure~\ref{f:lengthRatioScoreCutoff}, the score cutoff on the x-axis is the absolute value of the z-score. The absolute
value is used because this error detection system finds errors with both unusually high and unusually low z-scores. The y-axis
is the average number of Turkers (out of 5) who marked the proposed errors with scores above the
corresponding cutoff as real errors. Figure~\ref{f:transliterationScoreCutoff} shows similar information for the tied-field transliteration error
detection system. In Figure~\ref{f:transliterationScoreCutoff}, the score cutoff on the x-axis is the z-score. The z-score
is used because this error detection system finds errors with unusually high z-scores. The y-axis is the
average number of Turkers (out of 5) who marked the proposed errors with scores above the corresponding
score cutoff as real errors.

For systems that will be used for purposes of ranking proposed errors in order from most likely to least
likely, it is desirable that they predict errors with higher accuracy above a particular score threshold. 
This has positive implications for application settings where
users will go through errors in a sorted order from most likely to least likely.
Figure~\ref{f:lengthRatioScoreCutoff} and Figure~\ref{f:transliterationScoreCutoff} show that from this perspective the length ratio system produces better results. 
The length ratio system correctly predicts consistently with a cutoff over 5.5 standard deviations from the average. In contrast,
the transliteration system does not perform as well because it is unable to predict errors with as high a degree
of precision as the length ratio system at any score cutoff. A perhaps surprising result in Figure~\ref{f:transliterationScoreCutoff} is that the
transliteration system decreases in precision when the score cutoff increases past about 2.5 standard deviations
from the average. This result can be explained by the low number of proposed errors with very high scores with
the transliteration system. This can create the situation where there are too few data points to compute a
precision score reliably.

Table~\ref{t:overlap} displays the overlap of the errors most strongly proposed by the length ratio error detection system and the transliteration error detection 
system for varying cutoff sizes. This table allows us to evaluate the similarity between the two systems.
The two systems do not have a high degree of similarity; for example, only 4 of the top 100 anomalies detected
by the length ratio system also appear in the top 100 anomalies detected by the transliteration system. The
low levels of overlap indicate that the systems have complementary behavior. 
This complementary behavior can be leveraged to build improved hybrid systems. 

The four single-field methods can also be combined with each other and with the tied-field methods. 
By combining methods a hybrid system could be built that takes into account different perspectives on the data. 
There are many possible ways of combining methods to build a hybrid system, e.g., see \cite{bloodgood2012}.
In future work, it would be worthwhile to explore how to optimally combine the error detection methods to create an improved system. 

\begin{table}
\caption{Overlap of the anomalies most strongly proposed by the length ratio and the transliteration systems.}
\label{t:overlap}
\centering
\begin{tabular}{|c|c|c|}
\hline
Number of Proposed Errors & Number of Common Results & Percent \\
\hline
10 & 0 & 0\% \\ \hline
25 & 1 & 4\% \\ \hline
50 & 3 & 6\% \\ \hline
100 & 4 & 4\% \\ \hline
200 & 16 & 8\% \\ \hline
300 & 31 & 10\% \\ \hline
400 & 52 & 13\% \\ \hline
500 & 81 & 16\% \\ \hline
600 & 110 & 18\% \\ \hline
700 & 143 & 20\% \\ \hline
800 & 173 & 22\% \\ \hline
900 & 186 & 21\% \\ \hline
1000 & 227 & 23\% \\ \hline
1500 & 396 & 26\% \\ \hline
2000 & 563 & 28\% \\ \hline
\end{tabular}
\end{table}

We conducted an additional evaluation of our systems using the annotations of a language expert in the process of correcting errors in a Tamasheq dictionary containing 5968 lexical entries.
The language expert evaluated 175 anomalies proposed by the single-field text length system and the single-field uncommon character system for data 
in fields POS (part of speech) and MAIN (headword). 
The language expert marked each proposed anomaly as either a real error or not a real error. 
The results are in Table~\ref{t:languageExpertResults}. 
These results are further evidence that the error detection systems could substantially increase the efficiency with which domain experts can clean XML data.

\begin{table}
\caption{Language expert annotations on 175 anomalies proposed by the text length system and the uncommon character system for data in fields POS and MAIN.}
\label{t:languageExpertResults}
\centering
\begin{tabular}{|c|c|c|c|c|}
\hline
System & Field & Real Error & No Error & Total \\ \hline
Uncommon & POS & 16 & 1 & 17 \\ \cline{2-5}
Character System & MAIN & 4 & 10 & 14 \\ \hline
Text Length & POS & 110 & 3 & 113 \\ \cline{2-5}
System & MAIN & 30 & 1 & 31 \\ \hline
\end{tabular}
\end{table}

\section{Conclusion} \label{conclusion}

There is increasing interest in methods for computer-aided rapid data cleaning. 
We presented multiple methods for data cleaning of XML electronic dictionaries. 
The methods detect errors in the data content of the XML, unlike previous work that detected errors in the structure of the XML electronic dictionaries and ignored the content. 
The methods are based on different underlying sources and indicators of errors and have complementary behavior with each other and with previously developed methods.  
The methods can be classified into single-field methods and tied-field methods. 
Single-field methods detect anomalies on the basis of expectations inferred from content in the same single field type. 
Tied-field methods detect anomalies on the basis of expectations of content correspondences inferred from content in multiple related fields. 
Four single-field methods for error detection were presented that work by using expectations of word sequence information, 
expectations of character sequence information, expectations of length, and expectations of individual characters in particular fields. 
Two tied-field methods for error detection were presented that work by using expectations of length ratios and expectations of string correspondences via transliteration models. 

The precision of the error detection systems tends to correlate with the internal scores of the systems. 
This desirable behavior supports the scenario where a domain expert would go down a sorted list of
proposed errors from most likely to least likely. The performance of the different error detection systems varies
based on the XML fields and the dictionaries on which they are applied. Domain experts can choose to invoke error
detection system-field combinations with score cutoffs to suit their needs.

We evaluated these methods using the crowdsourcing platform Mechanical Turk and using expert annotations on multiple datasets. 
Sometimes the systems have overlapping behavior, detecting the same errors albeit through different views of the data. 
Often the systems have complementary behavior, which is promising for hybrid system construction in the future. 
We provided intuitive examples of the sorts of errors each of the systems can detect. 
In the evaluations, the systems are consistently helpful in increasing the efficiency with which data errors can be identified. 

\section*{Acknowledgment}

This material is based upon work supported, in whole or in part, with funding from the United States Government. Any opinions, findings and conclusions or recommendations expressed in this material are those of the author(s) and do not necessarily reflect the views of the University of Maryland, College Park and/or any agency or entity of the United States Government.

\bibliographystyle{IEEEtran}
\bibliography{paper}

\end{document}